\documentstyle[12pt,aasms4]{article}

\def\be{\begin{equation}}
\def\ee{\end{equation}}
\def\beq{\begin{eqnarray}}
\def\eeq{\end{eqnarray}}

\def\part{\partial}

\def\etal{{\it et al.\ }}

\begin{document}
\baselineskip=24pt
\begin{center}
{\Large\bf A Note on Power-Laws of Internet Topology}
\bigbreak
{\large\bf by}
\medbreak
{\large\bf Hongsong Chou } \\
{\it Harvard University, Cambridge, MA 02138}\\
{\it chou5@fas.harvard.edu}
\end{center}
\begin{abstract}
The three Power-Laws proposed by Faloutsos \etal(1999) are important
discoveries among many recent works on finding hidden rules in the
seemingly chaotic Internet topology. In this note, we want to point out
that the first two laws discovered by Faloutsos \etal(1999, hereafter,
{\it Faloutsos' Power Laws}) are in fact equivalent. That is, as long as any one of
them is true, the other can be derived from it, and {\it vice
versa}. Although these two laws are equivalent, they provide different
ways to measure the exponents of their corresponding power law
relations. We also show that these two measures will give equivalent
results, but with different error bars. We argue that for nodes of not
very large out-degree($\leq 32$ in our simulation), the first Faloutsos'
Power Law is superior to the second one in giving a better estimate of
the exponent, while for nodes of very large out-degree($> 32$) the
power law relation may not be present, at least for the relation
between the frequency of out-degree and node out-degree. 
\end{abstract}

\section{Introduction}
The past five years has been the golden time for the 30 year old Internet,
during which it experienced fascinating evolution, both exponential
growth in its traffic and endless expansion in its topology. Such
growth makes more thorough and rigorous analysis of the nature of
Internet traffic and topology an urgent task. It is also a very
difficult one. It was ever believed that the mathematical theories for
{\it circuit switching} telephone networks might be good enough for
analyzing the Internet traffic and topology. However, later it was found that
the Internet, as a {\it package switching} network, has very different
nature and successful mathematical theories for the Internet can be
quite different from those for telephone networks(see Willinger and
Paxson(1998) for more details). 

As the Internet grows at an astonishing speed, more and more high
quality data have been collected in recent years. These data make thorough
studies possible. The pioneering work by Leland \etal(1994) shows that
the traffic of Local Area Network(LAN) appears to be self-similar at different
scales. Discoveries of other self-similarities such as the one found in
Wide Area Network(WAN) by Paxson and Floyd(1995) make Internet
engineers and interested mathematicians contemplate that some special
power laws such as the heavy tail distribution might be the hidden
rules in Internet traffic(see Willinger and Paxson(1998) or Willinger,
Paxson and Taqqu(1998)).

The discovery of three Power Laws by Faloutsos, Faloutsos and
Faloutsos(1999) is one of the most recent work on the Internet
topology. They view the Internet as an undirected graph. For each node
in the graph, it has properties such as the out-degree. Faloutsos'
discovery is not just power laws for the large scale properties of
the Internet, but rather the relationships between nodes at different
scales, running from a host on a LAN to the range encompassed by the whole
Internet. Without a doubt, such discovery is important not only to our
understanding of the very nature of the rapidly growing Internet, but also
to any reasonable simulations of LANs or WANs or the whole Internet.

Yet, as we will point out in section 2 of this note, the first two Power Laws in
Faloutsos' discovery are not independent to each other. In fact, they
are equivalent so each one can be derived from the other. In section 3,
we
will go further to show that the data analysis in Faloutsos' work
toward discovering Power Law 1 is superior to the data analysis work
done for Power Law 2, simply because the former data analysis will give
more accurate estimate comparing to the later one. Conclusions are
summarized in the last section, section 4. 

\section{Equivalence Between the first and the second Faloutsos' Power
Laws}

Throughout this note, we adopt the notations used in Faloutsos'
work. The Internet is viewed as an undirected graph $G$, and the number
of nodes and the number of edges in $G$ are $N$ and $E$,
respectively. The out-degree of a node $v$, which is the number of edges
incident to the node, is denoted by $d_v$. Note that in $G$, different
nodes may have the same out-degree. That is, if we can group all the
nodes which have the same out-degree $d$ and index the group, then for
the nodes in the $l^{th}$ group, they have the same out-degree denoted
by $dl$. The number of nodes in this $l^{th}$ group, which gives the
frequency of appearances of out-degree $dl$ in $G$, is denoted by
$f_{dl}$. Sometimes we just write $f_d$ to denote the frequency of $d$
in $G$. This is because the out-degree $d$, which always starts from 1
throughout this note, can be used to index the groups of nodes of different
out-degrees, thus $f_d$ is the number of nodes in the $d^{th}$ group for
out-degree $d$. The {\it rank}, $r_{v}$, of a node $v$ which has an
out-degree $d$ is the global index of the node among all of the nodes
in the order of decreasing out-degree.

The first and second Faloutsos' Power Laws can be stated as
\be
d_v=C_1 r_{v}^R
\ee
and
\be
f_{d}=C_2 d^O
\ee
respectively. Here $C_1$ is a constant and can be determined by any given pair of
$d_v$ and $r_v$ measured from data collected from the Internet. $C_2$
is another constant and can be calculated from a pair of $f_{d}$ and
$d$. $R$ and $O$ are the two exponents of the Power Laws.

By definition, the frequency of out-degree $d$ in $G$, $f_d$, is
related to the ranks of those nodes which have out-degree $d$. Suppose
node $v_{d-1}$ is a node of out-degree $d-1$, and it is the last
indexed node with rank $r_v^{\prime}$ in the group consisting of nodes
which all have out-degree $d-1$. Further suppose that node $v_d$ is a node of
out-degree $d$, and it is the last indexed node, with rank $r_v$, in
the group consisting of nodes of out-degree $d$. Then $f_d$ is related
to $r_v^{\prime}$ and $r_v$ through the relation 
\be
f_{d} = r_v^{\prime} - r_v.
\ee
We may re-write relation (1) as
\be
r_v=\left ( \frac{1}{C_1} \right )^{\frac{1}{R}} d_v^{\frac{1}{R}}.
\ee
Note that the first order approximation to the right hand side of (3) is
in fact the first order derivative of the right hand side of (4) with
respect to $d$:
\be
f_d={r_v^{\prime} - r_v} \approx -\frac{1}{R}\left ( \frac{1}{C_1} \right
)^{\frac{1}{R}} d^{\frac{1}{R}-1}
\ee
From (4) to (5) we changed $d_v$ to $d$ because for all nodes of the group where
node $v_d$ is in, they all have the same out-degree $d$. Comparing (5) and (2), we have
\be
O \approx \frac{1}{R}-1 
\ee
and
\be
C_2 \approx -\frac{1}{R}\left ( \frac{1}{C_1} \right
)^{\frac{1}{R}}.
\ee
Thus we have derived the second Power Law from the first Power Law. To
derive the first Power Law from the second, we have to integrate
relation (2) from 1 to $d_v$ to get $r_v$, then compare the result with
relation (4). By doing this, we have
\be
R \approx \frac{1}{O+1}
\ee
and
\be
C_1 \approx \left (\frac{-O-1}{C_2} \right)^R.
\ee
(6), (7) and (8), (9) shows that whenever we have one of the two Power
Laws, exponent and the constant of the other one can be derived from
the given parameters. That is, in data analysis of the Internet
topology, once we have measured $r_v$ at different out-degrees and found
a power law relation with exponent $R$ between them, we do not need to measure
$f_d$ at different out-degrees because the power law relation between
$r_v$ and out-degree will guarantee the power law relation between $f_d$
and $d$ with an exponent $O \approx \frac{1}{R}-1$. In simulations,
samples generated according to Power Law 1 will follow Power Law 2
automatically, and {\it vice versa}.

In Table 1 and Table 2, we list the comparisons of
the derived parameters using above relations and the measured
parameters given in the work of Faloutsos'. From table 1 we find our
calculated exponent $O$ of Power Law 2 is quite close to the measured
one, except the last case, which is the Rout-95 dataset. The small
discrepancy shows that mere coincidence is not likely. For the
comparison of our calculated exponent $R$ and the measured $R$ in table
2, although the relative errors are larger than those in table 1, for
the first three cases they are still below 15\%.

\section{Better Way to Estimate Exponent}
When deriving relations (6) and (8) in above section, we assumed first
that one of the Power Laws must hold. In the derivation, we used
differentiation and integration, which can only be approximately
correct because the real datasets are discrete samples. Suppose at one
sampling position, such as an out-degree $d$, the measured rank is $r_v$,
and our calculated rank by integrating equation (2) is ${\hat r}_v$. We denote the
difference between $r_v$ and ${\hat r}_v$ by $\epsilon$, and call it an
{\it error term} for the estimation of rank at out-degree $d$. We
can define a similar error term, $\eta$, for the estimation of
frequency at out-degree $d$ with equation (5). The errors, both
$\epsilon$ and $\eta$, can be the measurement errors, the round-off
errors, or the errors due to the discrete nature of our sampling, and
in most cases, the combination of them all.

Non-zero $\epsilon$ will affect our estimation of $O$ made in table
(1). Non-zero $\eta$ will also affect our estimation of $R$ in table
(2), but in a different way from how $\epsilon$ affects estimating
$O$. We find that the relative errors for the first three cases in
table (1) are smaller than those in table (2). In other words, the
derivation of $O$ from $R$ by differentiating equation (4) gives closer
to measured results than the derivation of $R$ from $O$ by integrating
equation (2). This is because that if rank $r_v$ has error $\epsilon$,
then from equation (3) the error in $f_d$ will be of the order 
$\sim O(\epsilon)$. However, if $f_d$ has error $\eta$, then the error in
$r_v$, which can be obtained by integrating equation (2), is in fact an
accumulation of $\eta$ in the summation, which is $\sim O(n\eta)$ where
$n$ is the number of out-degrees used in the integration. Hence, even
though the two Power Laws are equivalent, deriving the second Power Law
from the first one will give better estimate, i.e., estimate with smaller
errors if $\epsilon \sim \eta$, of the second Power Law than the
estimate of the first Power Law derived from the second one.  

In Faloutsos' work, they applied linear regression to obtain the Power
Laws. We have shown above that Power Laws 1 and 2 are equivalent,
therefore the two linear regressions applied in Faloutsos'  work should
give the same answer.  That is, if we start from a Power Law relation
between rank $r$ and out-degree $d$, for example, $d = C_1 r^R$,  and
deduce the relation between frequency $f_d$ and $d$, the relation should also
be a power law. for example, $f_d = C_2 d^O$, and the exponent $O$
should be related to $R$ through (6). In Figure 1, the $\star$'s show the relation
\be
d = C_1 r^{-1.0}.
\ee
Note the logarithmic scales on both axes. There are 2000 data
generated, so the rank $r$ runs from 1 to 2000. The heavy solid line is
the linear fitting to the $\star$'s, with slope $-0.85$, instead of $-1$ as
we expect. This is due to the discretization of the data. The $\star$'s with
out-degree $d>1$ have a linear fitting of slope $-0.97$, which is shown
in Figure 1 as the dash line. Apparently the data of out-degree 1 have
large effect on the linear fitting.

In Figure 2, we plot the relation between $f_d$ and $d$ based on the
data($\star$'s) collected in Figure 1. A few data of frequency 1 and
out-degree $d>33$ are outliers and discarded in the fitting made in Figure
2, which is shown as a heavy solid line. The number of these discarded outliers
is 26, only 1.3\% of the total data. The slope of the linear fitting is
$-2.01$, which is what we expect because of the equivalence of the two
Power Laws, i.e., equation (6). 

\section{Discussions and Conclusions}
Given the definitions of frequency $f_d$ and rank $r$, it is not
surprising to see the equivalence of the first two Power Laws proposed by
Faloutsos \etal We have proved such equivalence and demonstrated the mutual
determination of these two relations, therefore it is not possible nor
necessary for any simulations to follow these two power relations
independently. However, as we have shown in section 3, determining the
power relation between frequency $f_d$ and out-degree $d$ by analyzing
the data of rank $r_v$ as a function of our-degree $d$, will give
estimates of smaller error comparing to the estimate made in reversed
order, i.e., the estimate of the power law relation between rank $r$
and out-degree $d$ by analyzing the data of frequency $f_d$ as a
function of out-degree $d$. For any set of data measured from the
Internet, they will follow the power law only {\it approximately}, not
always exactly, especially the nodes of very high out-degree and
frequency 1, or the nodes of out-degree 1, as we show in Figures (1) and
(2). In simulations, these nodes should be treated with special care.

If the probability density function for the appearance of nodes with out-degree
$s$ in the Internet is $\rho(s)$, then the average number of nodes
whose out-degrees run from $d_1$ to $d_2$ is
\be
\int_{d_1}^{d_2} s \rho(s) ds,
\ee
with which we can deduce the relation between frequency $f_d$ at out-degree
$d$ and the probability density function $\rho(s)$ as
\be
f_d \approx \int_{d-\Delta d}^{d+\Delta d} s \rho(s) ds \approx \rho(d)
d,
\ee
where we assume $2 \Delta d = 1$. If $f_d = C_2 d^O$, we have the
probability density function $\rho(d)$ as
\be
\rho(s) \approx C_2 d^{O-1}
\ee
which is a heavy tail distribution. The rank $r$ is related to the
function $\rho(d)$ through the integration
\be
r(d)=\int_1^d s \rho(s) ds
\ee
for $d>1$.

The Power Laws show the relations between nodes of different
out-degrees when the Internet is in steady state. In order to study
the dynamics of the Internet, it would be very interesting to inject
nodes of some specific out-degrees into the Internet, and follow the
temporal evolution of these nodes. By the time we inject nodes of some
specific out-degree, we alter the power law relationship between $f_d$
and $d$ by adding a spike-like disturbance(see Figure 3). If a steady
Internet does follow power laws, tracing the evolution of the
spike-like disturbance will tell us how the disturbance will be
propagated, or {\it cascaded}, toward higher out-degree and lower
out-degree directions(shown by the two arrows in Figure 3), the spike
being broadened at the same time(shown by the dash line in Figure
3). In real life, such spike-like disturbance could be due to the sharp
increase in the number of Internet users signing onto their
ISPs. Studies on the dynamic evolution of the Internet due to such
spike-like disturbances will be included in our future work.   
\acknowledgements

\clearpage
\begin{deluxetable}{ccccc}
\footnotesize
\tablecaption{Exponent O: measured and calculated with equation (6). \label{tbl-1}}
\tablewidth{0pt}
\tablehead{
\colhead{dataset} & \colhead{measured R} & \colhead{measured O} &
\colhead{calculated O } & \colhead{relative error}
}

\startdata
Int-11-97 &$-$0.81 &$-$2.15 &$-$2.23 &4\% \nl
Int-04-98 &$-$0.82 &$-$2.16 &$-$2.22 &4\% \nl
Int-12-98 &$-$0.74 &$-$2.20 &$-$2.35 &7\% \nl
Rout-95 &$-$0.48 &$-$2.48 &$-$3.08 &25\% \nl
\enddata

\end{deluxetable}

\clearpage
\begin{deluxetable}{ccccc}
\footnotesize
\tablecaption{Exponent R: measured and calculated with equation (8). \label{tbl-2}}
\tablewidth{0pt}
\tablehead{
\colhead{dataset} & \colhead{measured O} & \colhead{measured R} &
\colhead{calculated R } & \colhead{relative error}
}

\startdata
Int-11-97 &$-$2.15 &$-$0.81 &$-$0.87 &7.4\% \nl
Int-04-98 &$-$2.16 &$-$0.82 &$-$0.86 &5\% \nl
Int-12-98 &$-$2.20 &$-$0.74 &$-$0.83 &12\% \nl
Rout-95 &$-$2.48 &$-$0.48 &$-$0.68 &42\% \nl
\enddata

\end{deluxetable}

\clearpage

\begin{figure}[htbp]

\plotfiddle{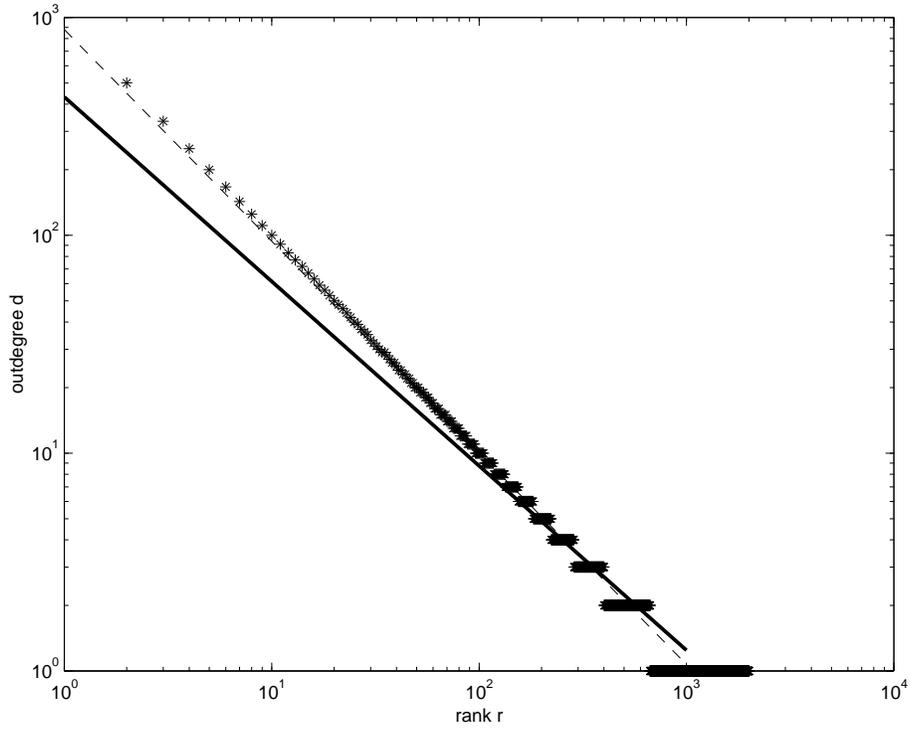}{6in}{0}{70}{70}{-210}{-25}

\caption{Out-degree $d$ {\it vs.} rank $r$. The $\star$'s are 2000 data
points obtained from the relation (10) in text. The heavy solid line of
slope $-0.85$ is the fitting to these data. Dash line is the fitting to
the data of out-degree greater than 1, with slope $-0.97$.
}
\end{figure}

\begin{figure}[htbp]

\plotfiddle{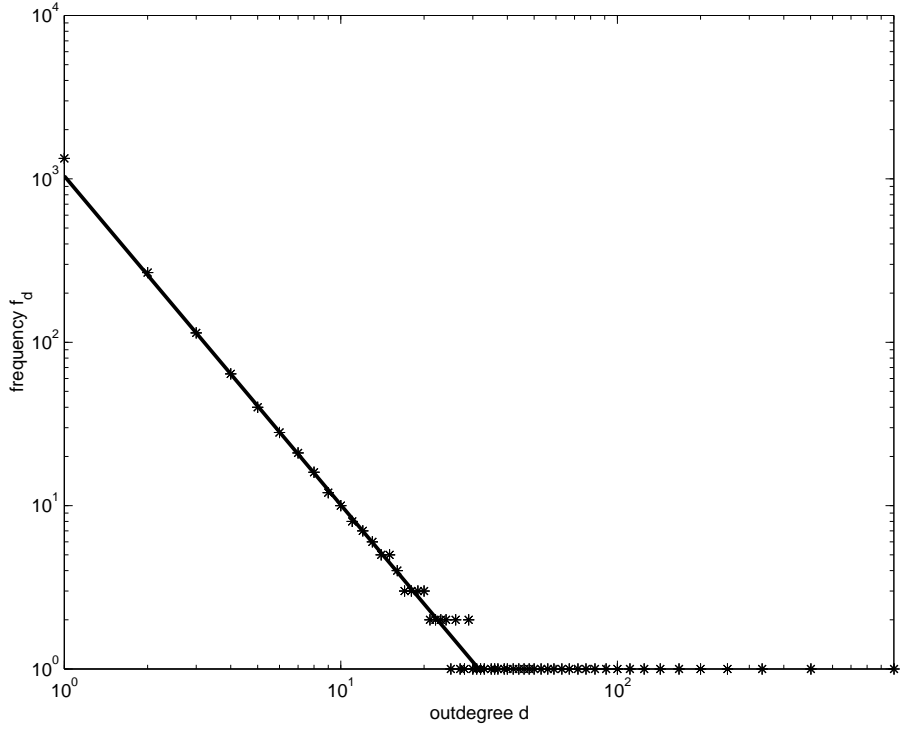}{6in}{0}{70}{70}{-210}{-25}

\caption{Frequency {\it vs.} out-degree $d$, following Fig. 1. The
$\star$'s are data calculated with relation (3) in text. The heavy solid line is the fitting to
data of out-degrees less than 33. The number of discarded data in the
linear fitting is 26, only 1.3\% of the total data.}
\end{figure}

\begin{figure}[htbp]

\plotfiddle{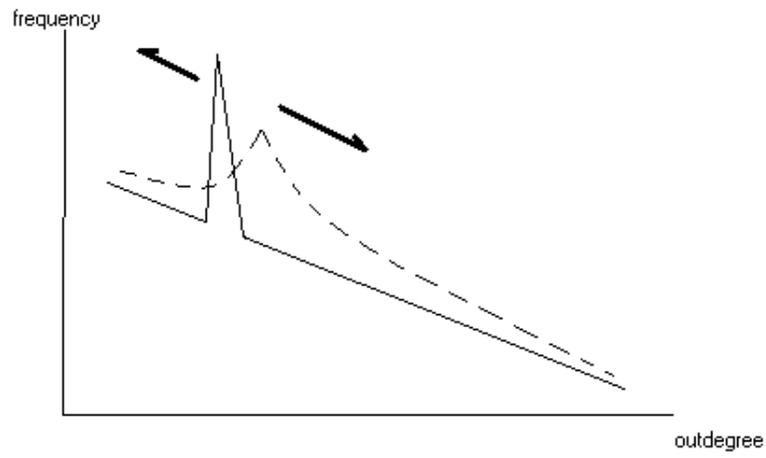}{6in}{0}{70}{70}{-210}{-25}

\figcaption{The cascade of spike-like disturbance of steady state
Internet toward large or small out-degree directions.}
\end{figure}

\end{document}